\documentclass[12pt]{article} 
\usepackage{amssymb,amsmath,latexsym}
\usepackage{graphicx}
\usepackage[margin=1.0in]{geometry}
\usepackage{color}
\usepackage[pdftex,colorlinks=true,urlcolor=cyan,linkcolor=blue,citecolor=blue]{hyperref}

\title{Even parity  potential for spacetimes with Gaussian matter source and its associated QNM spectra}
\author {\tt Souvik Pramanik,\,$^{(a)}$\thanks{\bf The two authors contributed equally to this work.}~\footnote{souvick.in@gmail.com} \hspace{4pt} Kumar Das\,$^{(b)}$\footnotemark[1]~\footnote{das.kumar582@gmail.com}  
\\ [10pt]
\small\em $^{(a)}$  Department of Physical Science, Indian Institute of Science Education and Research Kolkata, \\ 
\small\em Mohanpur Campus, West Bengal 741246, India. \\
\small\em $^{(b)}$ S. N. Bose National Centre For Basic Sciences, 
JD Block, Sector III, Salt Lake City, \\ 
\small\em Kolkata-700106, India. \\
}

\date{}

\begin{document}

\maketitle
    \begin{abstract}

We have obtained the general form of even mode potential function for the gravitational perturbation of a static spherically symmetric spacetime. Considering the minimal-spread Gaussian matter source, parameterized by a smearing length scale $\Theta$, the associated quasinormal mode (QNM) frequencies are computed numerically using the well known WKB formula. A comparative study on the accuracy of the results based on different approximation orders of the WKB formula indicates that the smearing parameter $\Theta$ has an upper bound, beyond which the spectra of odd and even modes deviate from their usual isospectral nature.

\end{abstract}


\vspace*{0.07cm}

\newpage

\section{Introduction}
\label{introduc}


The idea of Gravitational Waves (GWs) was explicitly put forwarded by Einstein in the theory of general relativity (GR), in 1916. From GR, we know that the acceleration of masses generates time-dependent gravitational fields. These fields propagate away from their sources as 'ripples' in the fabric of space-time at the speed of light. Such propagating waves are called gravitational waves. The first direct detection of a gravitational wave signal (GW150914) coming from a binary systems of massive stellar black holes (BH), was announced by the LIGO and VIRGO Collaboration \cite{Abbott:2016blz}. Thereafter, a series of such events have been recorded \cite{Abbott:2016nmj,Abbott:2017vtc,TheLIGOScientific:2017qsa}. These results motivate us to explore a whole new perspective about the universe, where much exciting physics are awaiting to be deciphered. The physical processes that are responsible for the production of astrophysical GWs involve strong curvature of the spacetime geometry, where only gravity plays the dominant role over other known interactions of nature. In the absence of a mathematical form of the metric describing BH binaries, the studies within a linearized approximation to Einstein's equations aided with full-blown numerical simulations, are found to be consistent with the predictions of GR \cite{Will:2014kxa}. Later works have shown that there are significant deviations too \cite{kon}. Nevertheless, at present tremendous efforts are continuously being made to understand and predict the GW signals coming from all possible sources of radiation. Such predictions will then help us to construct GW templates for extracting information about the GW source. Hence, it is important to recognize gravitational radiation from other common objects that can also become relevant for new generation GW antennas. 

 Once a BH is perturbed, its perturbation evolves in three different stages. First, there is an initial outburst of radiation lasting for a short time. After that, a long period of damped oscillations takes place at the intermediate stage when the system loses energy by gravitational radiation. The frequency of these oscillations are usually known as the Quasinormal modes (QNMs) \cite{Vishveshwara:1970zz,Macedo:2016wgh}.
 At very late times these modes are suppressed by a power-law fall off (for a thorough account on the QNMs see the reviews \cite{Konoplya:2011qq,Cardoso:2003pj,Berti:2009kk,Berti:2005ys,Pretorius:2005gq})

In this work, we have studied the QNMs of gravitational perturbations from spherically symmetric spacetimes having a Gaussian distributed  mass profile. 
Originally, this type of matter distribution was tacitly exploited by Nicolini \emph{et.al.} to successfully cure the singularity problem of the standard Schwarschild BH \cite{Nicolini:2005vd}. Also the associated smearing length scale was supposed to have an interpretation in the context of quantum gravity.
 Thereafter, an exhaustive treatment of these generalized BHs have been made in various contexts (see for details \cite{Nicolini:2008aj}, \cite{Pramanik:2015eka}, etc.). From an observational point of view, the relevance of such a Gaussian mass profile was discussed in our earlier work 
\cite{Das:2018fzc}. For the present work, we focus on even parity perturbations of this smeared geometry where the smearing length scale is denoted by the parameter $\Theta$.

 The organization of our paper is as follows --- in Sec.~\ref{pert_space}  we have briefly reviewed the gravitational perturbations of a static spherically symmetric spacetime. For the even parity components of this gravitational perturbation, a general form for the potential function was derived. Thereafter, in Sec.~\ref{wkbkono} we review the basic aspects of the QNM and the standard procedure to estimate them using the WKB approximation formula. In Sec.~\ref{gwgauss}, we found the form of the even mode potential function in the presence of a Gaussian distributed matter source. Soon after this, in Sec.~\ref{numer_qnm} we numerically estimate the QNM frequencies for these modes using the 6th order WKB formula  and compared the results due to different orders in the approximation in WKB series. Here, we also comment on the isospectrality between odd and even parity perturbation that holds upto a limitting value of the smearing parameter. Finally, in Sec.~\ref{conl} we concluded our results.


\section{Perturbation equations and general form of the even parity potential} 
\label{pert_space}


\vspace{0.2cm}

It is known that black holes (BH) have characteristic oscillation frequencies. They arise when we deal with the evolution of some perturbation in the BH spacetime, or in BH-BH collision processes. The frequency and damping of these oscillations are completely independent of the particular initial configuration, that caused the excitation of such vibrations and depend only on the parameters characterizing the BH (\emph{e.g.} in the Schwarzschild case it is its mass). Such oscillations have been termed ``quasinormal modes (QNM)''and the associated complex frequencies ``quasinormal mode frequencies''. The real part of a QNM frequency corresponds to the actual oscillation frequency, whereas the absolute value of the imaginary part represents the rate at which each mode damps or grows. One usually studies QNMs, specially the lowest lying modes, in order to get a better understanding of the properties of the gravitational wave signal. As these QNMs are independent of the details of initial perturbations, thus they allow us probe the black hole mass, electric charge and angular momentum \cite{Echeverria:1989hg}, through their characteristic waveform.

Much effort has been spent in developing powerful methods, both analytical and numerical, to calculate the QNMs and their associated frequencies \cite{Leaver:1985ax,chandra,Schutz:1985zz,Iyer:1986np,Konoplya:2010kv}. The main interest in these studies is in the application to the analysis of the data from the GWs to be detected by the forthcoming GW detectors.

In our earlier work \cite{Das:2018fzc}, we have found the QNM frequencies for odd parity perturbations from a spherically symmetric spacetime with smeared matter source. In this paper, we are interested in the QNMs for even party perturbations of the same geometry. 

To begin with, we consider the evolution of gravitational perturbations of a stationary spherically symmetric spacetime. The metric of such spacetime is given by
\begin{align}
ds^2 = -f(r) dt^2 + \frac{dr^2}{f(r)} + r^2 d\Omega_{2}^2
\label{sphe_me}
\end{align}
where, $d\Omega_{2}^2 = d\theta^2 +\sin^2\theta d\phi^2$ is the line element of a 2-sphere. For a Schwarzschild BH, $f(r)=1-\frac{2M}{r}$, $M$ being the BH mass.

For first order gravitational perturbations, the metric function can be written as 
(for a review on the BH perturbations in the context of possibility to observe QNM ringing with the help of GW detectors see \cite{Nollert:1999ji,Kokkotas:1999bd}), 
\begin{align}
 g_{\mu\nu}(x^{\alpha}) = \tilde{ g}_{\mu\nu}(x^{\alpha}) +  h_{\mu\nu}(x^{\alpha})  \qquad\text{where,} \qquad |h_{\mu\nu}| << |\tilde{ g}_{\mu\nu}|.
\label{nw_mec}
\end{align}
where $\tilde{ g}_{\mu\nu}$ is the background metric given by eqn.~\eqref{sphe_me} and $h_{\mu\nu}$  is a small perturbation.  
 Since the perturbation is assumed to be small (\emph{i.e.} we neglect terms like $\mathcal{O}(h^2)$ and so on ), all covariant derivative ($\nabla_{\mu}$) can be taken with respect to the background metric $\tilde{g}_{\mu\nu}$.

The Einstein field equation for this system will be 
\begin{align}
R_{\mu\nu}(\tilde g + h) = R_{\mu\nu}(\tilde{g}) + \delta R_{\mu\nu}(h) = 0
\label{ei_eqs}
\end{align}
where $R_{\mu\nu}(\tilde g + h)$ is the Ricci tensor computed from the total metric of eqn.\eqref{sphe_me} and $R_{\mu\nu}(g)$ is the same obtained from $\tilde{g}_{\mu\nu}$ which we know will vanish. Now $\delta R_{\mu\nu}(h)$ can be expressed as $\delta R_{\mu\nu}(h) =  - \nabla_{\nu}\delta\Gamma_{\mu\beta}^{\beta} + \nabla_{\beta}\delta\Gamma_{\mu\nu}^{\beta}$, where the variation of affine connection is
\begin{align}
 \delta\Gamma^{\alpha}_{\mu\nu} =  \frac{\tilde{ g}^{\alpha\beta}}{2}\big(\nabla_{\nu}h_{\mu\beta} + \nabla_{\mu}h_{\nu\beta} - \nabla_{\beta}h_{\mu\nu} \big)
\end{align}
Therefore, the field eqn.~\eqref{ei_eqs} becomes $\delta R_{\mu\nu}(h) = 0$. Upon employing gauge freedom in the above equation, we get the following second order differential equation for the perturbations  
\begin{align}
\Box h_{\mu\nu} - 2\tilde{R}^{\kappa}_{\,\sigma\mu\nu}h_{\kappa}^{\,\sigma} = 0
\end{align}
with the TT (transverse traceless) gauge, where 
\begin{align}
\nabla^{\mu}h_{\mu\nu}=0 \quad \text{and} \quad h_{~\mu}^{\mu} = \tilde{ g}^{\mu\nu}h_{\mu\nu} = h = 0.
\end{align} 
Any arbitrary perturbations of a spherically symmetric black hole in (3+1) dimensions can be decomposed in terms of normal modes. For any given value of the angular momentum $(L)$, associated with these modes, there are two distinct classes of perturbations --- odd and even subjected to parities $(-1)^{L+1}$ and $(-1)^{L}$, respectively \cite{chandra,Regge:1957td,Zerilli:1971wd}. For odd parity modes ($h_{\mu\nu}^{\text{odd}}$) in the presence of smeared matter sources, the associated QNM spectra was found in \cite{Das:2018fzc}.
Here we shall proceed in dealing with even parity perturbations ($h_{\mu\nu}^{\text{even}}$). Perturbations of this type are also known as \emph{polar} perturbations.
The mathematical analysis for polar perturbations is very similar to that followed for axial perturbation.
To simplify the elements of $h_{\mu\nu}^{\text{even}}$, one can use the invariance under infinitesimal coordinate transformations \emph{i.e.} $x^{\alpha\prime}\to x^{\alpha} +\xi^{\alpha}(x)$. This gauge freedom allows to simplify perturbation equations by eliminating all the highest derivatives in the angles $(\theta,\phi)$(see \cite{Regge:1957td, Rezzolla:2003ua} for details).
After this simplification, polar metric perturbations are described by three unknown functions $K(t,r),H(t,r)$ and $H_1(t,r)$. For the metric of eqn.~\eqref{sphe_me}, the system of equations governing these functions are 
\begin{eqnarray}
\frac{dK}{dr} = \frac{f'(r)K[r]}{2f(r)} - \frac{K[r]}{r} + \frac{H(r)}{r} + \frac{iL(L+1)H_1(r)}{2 r^2 \omega} 
\label{e1}
\end{eqnarray}
\begin{eqnarray} 
\frac{dH_1}{dr} = -\frac{i \omega  K[r]}{f(r)} - \frac{H_1(r)f'(r)}{f(r)} - \frac{i\omega H(r)}{f(r)} 
\label{e2} 
\end{eqnarray}
\begin{eqnarray}
\frac{dH}{dr} = \frac{f'(r) K[r]}{2 f(r)} - \frac{K[r]}{r} - \frac{H(r)f'(r)}{f(r)} - \frac{i\omega H_1(r)}{f(r)} 
                     + \frac{H(r)}{r}+\frac{i\,L(L+1)H_1(r)}{2 r^2 \omega}
\label{e3}
\end{eqnarray}  
where the time dependence is extracted for $K$ as $K(t,r)\sim e^{i\omega t} K(r)$ and similarly for other functions as well.

 Now for even parity modes, there are actually six equations governing the unknown functions $K,H$, and $H_1$. Out of them the three first order differential eqns.~\eqref{e1} --- \eqref{e3} are sufficient to determine a solution, as they are independent. It turns out that the rest of the three second order differential equations determine a constraint relation along with eqns.~\eqref{e1} --- \eqref{e3} \cite{Regge:1957td,Zerilli:1971wd}. The constraint in this case is as follows
\begin{eqnarray}
H[r] &=&  \frac{2 r \omega  f(r) \left[f'(r)-\frac{2 r \omega ^2}{f(r)}\right] - r^2 \omega  f'(r)^2 + 2 \big(L (L+1)-2\big) \omega  f(r)}
{2 \omega  f(r) \left[r
	f'(r)+f(r) \left(\frac{L (L+1)}{f(r)}-2\right)\right]} K[r] \nonumber\\
&& + ~ \frac{i \big(4 r \omega ^2-L (L+1) f'(r)\big)}{2 \omega  \left[r f'(r)+f(r) \left(\frac{L
		(L+1)}{f(r)}-2\right)\right]} H_1(r) 
\label{constrn}
\end{eqnarray}

Following Zerili \cite{Zerilli:1971wd}, we rescale the function $H1$ as $R(r) = \frac{H_1(r)}{\omega}$. Now we perform the transformation   
\begin{equation}
K\to f_1(r) \ \hat{K}(r) + f_2(r) \ \hat{R}(r) , \ \ R \to f_3(r) \ \hat{K}(r) + f_4(r) \ \hat{R}(r).
\label{func_trans}
\end{equation}
Here $\hat{K}, \hat{R}$ are dynamical variables and $f_1,f_2,f_3,f_4$ are arbitrary functions of $r$. Now we let $\hat K(r(r_{\star}))$ and $\hat R(r(r_{\star}))$ as functions of the so called tortoise coordinate $r_\star$, which is defined as 
\begin{align}
\frac{dr}{dr_\star} = f(r).
\label{co_rst}
\end{align}

Then using the defination of $R(r)$, the replacements \eqref{func_trans} and the constraint eqn.~\eqref{constrn}, the above system of eqns.~\eqref{e1} --- \eqref{e3} can be reduced to provide
\begin{eqnarray}
\frac{d\hat{K}(r(r_{\star}))}{d r_\star} &=&  \hat{R}(r(r_{\star})) - \tau_1(r) \hat{K} (r(r_{\star})), \\
\frac{d\hat{R}(r(r_{\star}))}{d r_\star} &=& \big(\tau_2(r) - \omega^2 \big)\hat{K}(r(r_{\star})),
\end{eqnarray}
where the algebraic functions $\tau_1(r)$ and $\tau_2(r)$ are given by
\begin{align}
\tau_1(r) = \frac{f(r) \left(r^2 f''(r)-2 f(r)+2\right) }{r(r f'(r)-2 f(r)+2 \lambda + 2 )}, \ \ 
\tau_2(r) = \frac{f(r)g(r)}{r^2 \left(r f'(r)-2 f(r)+2 \lambda +2\right)^2} \nonumber
\end{align}
with
\begin{align}
g(r) & = 2\big(r^2 f''(r)+3 r f'(r)+4 \lambda +6 \big)f(r)^2  -  \big\{2 r f'(r) (2 r f'(r)+5 \lambda +6 ) \nonumber \\
     &  \qquad  + r^2 (r f'(r)+4 )f''(r) + 12 (\lambda +1)^2 \big\}f(r) + \big(r f'(r)+2 \lambda +2\big)  \nonumber \\
     &  ~~~~ \qquad \big[(r^2 f''(r)+4 \lambda +2)(\lambda +1) + r^2 f'(r)^2 + 2 r f'(r)\big] -4 f(r)^3 \nonumber
\end{align}
and $\lambda = \frac 1 2 (L-1)(L+2)$. 
These two equations can be combined to provide a second order differential equation for the dynamical variable $\hat K(r_{\star}(r))$ as
\begin{align}
 \frac{d^2\hat{K}(r(r_\star))}{dr_{\star}^2}  + \tau_1(r)
  \frac{d\hat{K}(r(r_\star))}{dr_\star} + 
\bigg(\omega^2 -  \frac{\tau_2(r)}{g(r)}\big[g(r)-h(r)\big]  \bigg)\hat{K}(r(r_{\star})) = 0
\end{align}
where,
\begin{align} 
h(r) & =  r f'(r)\big(r f'(r) + 2 \lambda + 2\big)\big(r^2 f''(r) + 2\big) + f(r)^2\big[8 r f'(r) - 2 r^3 f'''(r) 
\nonumber \\
 &  \qquad ~~ + 4(\lambda +2)\big] + f(r)\big[r\big\{r^2\big(rf'(r)+2\lambda +2)f'''(r) + 2 r \lambda f''(r)  \nonumber \\
 & \qquad ~~ -r^3 f''(r)^2  - 4 f'(r)(3 + 2 \lambda + r f'(r)) \big\}-4 (1 + \lambda) \big] - 4 f(r)^3 \nonumber
\end{align}
Finally, the second term of the above equation can be eliminated by means of some transformation to yield a Schrodinger-type equation for the even mode gravitational perturbation and it is 
\begin{equation}
\frac{d^2\hat{K}}{d r_{\star}^2}  +  \big[\omega^2 - V_{\text{even}}(r)\big] \hat{K} = 0
\end{equation}
where $V_{\text{even}}(r)$ is the potential for even parity perturbation which has the following form

\begin{align}
V_{\text{even}}(r) &= \frac{f(r)}{4 r^2\big(r f'(r)-2 f(r)+2 \lambda + 2\big)^2} 
\bigg[\mathcal A_1(r) + \mathcal A_2(r) f(r) \nonumber \\
&  \qquad ~~~~ + \mathcal A_3(r)f^2(r) + \mathcal A_4(r)f^3(r) + \mathcal A_5(r)f^4(r)+ 4f^5(r) \bigg]
\label{even_poln}
\end{align}
where,
\begin{align}
\mathcal A_1(r) &= - 4(r f'(r) + 2\lambda + 2)\big[r\big(f'(r)(r f'(r) + 2) + r(\lambda + 1)f''(r)\big)
\nonumber \\
& \hspace{1.5cm} + 4\lambda^2 + 6\lambda + 2\big] \nonumber\\
\mathcal A_2(r) &= 8\lambda (11 + 6\lambda)f(r) + 2 r \big[6rf'(r)^2 +f'(r)(4r^2f''(r)+r^3f'''(r)
\nonumber \\
& ~~ +16\lambda + 20) + r\big(f''(r)(2\lambda-r^2f''(r)+8) + 2r(\lambda+1)f'''(r)\big)\big] \nonumber\\
\mathcal A_3(r) &= -2\big( 8rf'(r) + 8 r^2 f''(r) +r^4 f''(r)^2 +2 r^3 f'''(r)+12\lambda + 20 \big) \nonumber\\
\mathcal A_4(r) &= r^2 f''(r)(r^2 f''(r) + 12) + 28 \nonumber\\
\mathcal A_5(r) &= 4 + r^2 f''(r) \nonumber
\end{align}
This is the most general form for the effective potential of even parity gravitational perturbation from a spherically symmetric spacetime metric, given by (\ref{sphe_me}). Till now, in the literature this form was missing. So, this is an interesting thing we have evaluated here. Later on, we will use this general form to compute the even mode potential due to a smeared mass distribution. It should be noted that for a  Schwarzschild BH, plugging $f(r)=1-\frac{2M}{r}$ in eqn.~\eqref{even_poln} reproduces the famous Zerili potential ($V_{ZW}$)
\begin{align}
V_{ZW}(r) = \frac{2 (r-2 M)(9 M^3+9 \lambda M^2 r+3 \lambda^2 Mr^2+\lambda^2(\lambda +1) r^3)}{r^4 (3 M+\lambda r)^2}
\label{zeri_potl}
\end{align}
The solutions of eqn.\eqref{even_poln} define the QNMs of the black hole with QNM mode frequencies $\omega$. In the following subsection we describe in brief how to compute this frequency.


\subsection{Computing the QNM spectrum with $6$'th order in WKB approximation}
\label{wkbkono}
\vspace{0.2cm}

The first semi-analytic method for computing the QNM frequencies was suggested by Ferrari and Mashoon \cite{Ferrari:1984zz}. They compute the QNMs using their connection with bound states of the inverted BH effective potentials\footnote{The effective BH potential in Mashoon approach is the P$\ddot{\text o}$schel-Teller potential. See \cite{Das:2018fzc} where we have also computed the QNM spectrum for odd party modes due to spacetimes with smeared matter sources using this method and compared the accuracy of the resulting spectrum with that obtained by 3rd and 6th order WKB method}. For asymptotically flat spacetime, QNMs was numerically computed by Chandrasekhar and De Wittler \cite{chandra}. For the present work the spacetime under investigation is a QG-inspired spherically symmetric BH space-time \cite{Nicolini:2008aj}. In this case, a stable numerical method for getting QNMs does not exist, so far. So, we have to rely on certain approximation schemes of which WKB formula is found to be of wide use in the literature (see \cite{Konoplya:2011qq} for a detailed review on the aspects of various methods).
To estimate the QNM frequencies in this work, we will use the 6th order WKB formula. The WKB technique was first applied by Schutz and Will \cite{Schutz:1985zz} for finding QNMs of BHs. Later, the procedure is extended to the 3rd order beyond the eikonal approximation by Iyer  and subsequently to 6th order by Konoplya \cite{Iyer:1986np,Konoplya:2003ii}. The technical details of this approach are described in \cite{Konoplya:2003ii,Konoplya:2004ip,Konoplya:2002ky} (see also \cite{Kokkotas:1993ef,Andersson:1996xw,Onozawa:1995vu} for other uses of WKB method). 

The evolution of BH perturbations of a spherically symmetric spacetime are generically described by a Schr$\ddot{\text o}$dinger like wave equation 
\begin{align}
-\frac{d^2\Psi(x)}{dx^2} + V(x)\Psi(x) = \omega^2\Psi(x), 
\end{align}
where the functional form of the potential $V(x)$ depends on the specific field under consideration. Typically $V(x)$ approaches a constant at $x\to\pm\infty$ and at some intermediate value $x_0$, it rises to a maximum. Here $\Psi(x)$ represents the function describing perturbations of different kinds associated with a definite parity in the background static spacetime. For the metric of a spherically symmetric spacetime the co-ordinate $x$ plays the role of the tortoise coordinate $r_{\star}$ and in our work this potential function is due to the even parity gravitational perturbation given by expression \eqref{even_poln}. 
Hence, with this identification the problem now reduces to a problem of scattering near the pick of the potential barrier in quantum mechanics. QNMs are basically the eigenvalues of this Schr$\ddot{\text o}$dinger type equation. 

To find the eigenvalues in WKB approach one splits the potential into three regions --- the barrier regime, the event horizon and the spatial inifinity. The asymptotic WKB solutions for $\Psi(x)$ at spatial infinity and at event horizon are then  matched with the Taylor expanded wavefunction near the top of the potential barrier through the turning points \cite{Konoplya:2011qq,Konoplya:2010kv}. Finally, the resulting eigenvalue spectrum is given by the following formula \cite{Konoplya:2010kv,Konoplya:2003ii} 

\begin{align}
\omega^2 = V_0 - i \sqrt{-V_2} \bigg(n+\frac1 2 \bigg) + \sum_{i=2}^6 A_i  \qquad n=0,1,2,...
\label{6wkb}
\end{align}
where $A_i$'s represent i-th order correction in the WKB formula \emph{e.g.},
\begin{align}
A_2 = ~&(-11 V_{3}^2 + 9 V_2 V_4 - 30 V_{3}^2 n + 18 V_2 V_4 n - 30 V_{3}^2 n^2 + 18 V_2 V_4 n^2)/(144 V_{2}^2) \\
\frac{i A_3}{\sqrt{-2 V_2}} = ~& (-155 V_{3}^4 + 342 V_2 V_{3}^2 V_4 - 63 V_{2}^2 V_{4}^2 - 156 V_{2}^2 V_3 V_5 + 36 V_{2}^3 V_6 - 545 V_{3}^4 n \nonumber \\ 
&  + 1134 V_2 V_{3}^2 V_4 n - 177 V_{2}^2 V_{4}^2 n - 480 V_{2}^2 V_3 V_5 n + 96 V_{2}^3 V_6 n - 705 V_{3}^4 n^2 + 
\nonumber \\
& 1350 V_2 V_{3}^2 V_4 n^2 - 153 V_{2}^2 V_{4}^2 n^2 - 504 V_{2}^2 V_3 V_5 n^2 + 72 V_{2}^3 V_6 n^2 - 470 V_{3}^4 n^3 
\nonumber \\
& + 900 V_2 V_{3}^2 V_4 n^3 - 102 V_{2}^2 V_{4}^2 n^3 - 336 V_{2}^2 V_3 V_5 n^3 + 48 V_{2}^3 V_6 n^3)/(6912 V_{2}^5) .
\end{align} 
The full expressions for other correction terms $A_4,A_5, A_6$ are given in \cite{Konoplya:2003ii}. In the above expression, $V_0 (\tilde{r}_{\star})$ is the value of the effective potential in its maximum ($r=\tilde{r}_{\star})$ and $V_i(\tilde{r}_{\star})$ is the $i$-th derivative of the potential in tortoise coordinate in the maximum. We will employ this formula in the next section for calculating the QNMs for even mode gravitational perturbations for a spacetime with diffused mass distribution.


\section{Even mode potential for Gaussian matter distribution}
\label{gwgauss}

Now we consider the same spherically symmetric metric of eqn.~\eqref{sphe_me} but this time with a minimal-spread Gaussian profile instead of the conventional delta function type source. Except a motivation from the quantum gravity perspective, this effective description through Gaussian matter sources also has an astrophysical interest (see also \cite{hogg:1965hg,Press:1973iz} for details). In our earlier attempt, we computed the QNMs for odd parity perturbation of this system 
\cite{Das:2018fzc}. This work aims at obtaining the QNM frequencies for even mode perturbation of the corresponding system employing the powerful method of WKB approximation.


The spacetime of a spherically symmetric Gaussian distribution of mass is described by the following metric \cite{Nicolini:2008aj}

\begin{align}
ds^2 = -\bigg(1-\frac{4M}{r\sqrt\pi}\gamma(3/2, {r^2}/{4\Theta})\bigg)dt^2+\bigg(1-\frac{4M}{r\sqrt\pi}\gamma(3/2, r^2/{4\Theta}) \bigg)^{-1}dr^2 
+ r^2 d\Omega_{2}^2   
\label{qg_sphe}
\end{align} 
where $M$ is the total mass of the source diffused through out a region of minimal length scale $\sqrt{\Theta}$. Comparing the form of this metric with that given in eqn.~\eqref{sphe_me}, we find that
\begin{align}
f(r) = 1-\frac{4M}{r\sqrt\pi}\gamma(3/2, {r^2}/{4\Theta}) ,
\end{align}
where, $\gamma(3/2, r^2/{4\Theta})=\int_{0}^{r^2/{4\Theta}}\sqrt t\,e^{t}\,dt$ being the lower incomplete Gamma function. If the sprade of matter distribution is small compared to the size of object, then we can consider the limit $r^2>>4\Theta$, so that $f(r)$ becomes \cite{Nicolini:2008aj}
\begin{align}
f(r)= 1- \frac{2M}{r} + \frac{2M}{\sqrt{\pi\Theta}}e^{-r^2/{4\Theta}}.
\label{new_sph}
\end{align}

Likewise in Sec.~\ref{pert_space}, we perturb this metric and assume that perturbation is small with respect to the background. The Einstein equation for this perturbation will again be described by the same eqn.~\eqref{ei_eqs}. Our goal is to get the the form of the effective potential for even parity component of the perturbation. To do that, we plug the function $f(r)$ from eqn.~\eqref{new_sph} into the expression of the even parity potential derived earlier in eqn.~\eqref{even_poln}. This gives
\begin{align}
V_{even}^{\Theta}(r) &= V_{ZW}(r) + \frac{2M}{\sqrt{\pi\Theta}}e^{-r^2/{4\Theta}}\,\bigg[\mathcal C_0  + \sum_{i=1}^{9} C_ir^i 
 + \lambda^2 r^{10} \bigg]  \times \nonumber \\
& \hspace{3in} ~~~~ \frac{1}{32(3M+r\lambda)^3r^3\Theta^3}
\label{even_th}
\end{align} 
where, $\mathcal{C}_0$ and the various coefficients $C_{i}$ are 
\begin{align}
\mathcal C_0 &= 2880 M^4 \Theta^3 \\
\mathcal C_1 &= 576 (6\lambda - 1)M^3 \Theta^3 \\
\mathcal C_2 &= 96 \big(12 M^2 + \lambda (18 \lambda - 1)\Theta \big) M^2 \Theta^2 \\
\mathcal C_3 &= 16 \big[36 M^2 (\lambda - 1) + \lambda \big\{ 4\lambda(5\lambda - 3) - 15 \big\}\Theta\big] M \Theta^2  \\
\mathcal C_4 &= 8 \big[-63 M^4 + 3  \lambda ( 6 \lambda - 7) M^2 \Theta
 + 2 \lambda^2 (3 + 6 \lambda + 4\lambda^2)\Theta^2 \big]\Theta   \\ 
\mathcal C_5 &= 4 [36 M^2 (3 - 2 \lambda) + \lambda (15 + 24 \lambda + 4 \lambda^2)\Theta \big] M \Theta  \\
\mathcal C_6 &= 36 M^4 + 4 (3 + 2 \lambda)\lambda^2 \Theta^2 - 2 \big[45 + 2 \lambda(16 \lambda - 51)\big]M^2 \Theta \\
\mathcal C_7 &= 2 M (-3 + 2 \lambda) (6 M^2 + (5 - 2 \lambda)\lambda \Theta ) \\
\mathcal C_8 &= 4 \lambda^3 \Theta  + M^2 (9 - 24 \lambda + 4 \lambda^2) \\
\mathcal C_9 &= 2 M (3 - 2 \lambda) \lambda 
\end{align}

This is the form of the effective potential for even mode perturbations due to a Gaussian matter source. Our next task is to find the extremum of this potential. Let $r_{m}$ be the value of $r$ at which the potential $V_{even}^{\Theta}(r)$ has an extremum. In this case, we consider that $r_{m}$ has undergone a perturbative shift as,
\begin{equation}
r_{m} \to  r_0 + \frac{2M}{\sqrt{\pi\Theta}}\,e^{-r_{0}^2/{4\Theta}}\,r^{\prime},
\end{equation}
where, $r_0$ is the minimum of the Zerili potential of eqn.~\eqref{zeri_potl} obtained from $dV_{ZW}/dr=0$. Therefore, to the first order in the perturbation, $r^{\prime}$ is obtained from taking the first derivative of eqn.~\eqref{even_th} and it is 
\begin{align}
r_{m} = r_0 - \frac{2M}{\sqrt{\pi\Theta}}e^{-r_{0}^2/{4\Theta}}\frac{r_{0}^2}{\mathcal{S}(r)} \times \sum_{i=1}^{12}\mathcal{D}_{i}\,r_{0}^i
\end{align}
where, $\mathcal{D}_i$'s are given by 
\begin{align}
\mathcal D_0 &= 414720 M^5 \Theta^4 \\
\mathcal D_1 &= 27648 \big(11 L (L+1) -24 \big) M^4 \Theta^4 \\
\mathcal D_2 &= 2304 \big[54 M^2 + (L - 1) (L + 2)\big(39 L (L + 1)-89\big) \Theta\big] M^3 \Theta^3  \\
\mathcal D_3 &= 768 \big[9 \big(13 L (L + 1) -28\big) M^2 +  2 (L^2 +L-2)^2 \big(9 L (L + 1)-19 \big) \Theta\big] M^2 \Theta^3  \\
\mathcal D_4 &= 192 \big[270 M^4 + 3 (L-1) (L + 2) \big(39 L (L + 1)-101\big) M^2 \Theta  
\nonumber \\
& ~~~~ + (L^2 +L -2)^2 \big\{16 + L (L + 1)\big(4 L (L + 1)-25\big)\big\} \Theta^2\big] M \Theta^2   \\ 
\mathcal D_5 &= 64 \Theta^2 \big[54 \big(5 L (L + 1)-26\big) M^4 + 6 (L^2 +L-2)^2 \big(9 L (L + 1)-19\big) M^2 \Theta 
\nonumber \\
& ~~~~ + (L^2 +L-2)^3 (L^4 + 2L^3 - L + 1) \Theta^2\big]  \\
\mathcal D_6 &= 48 \big[-360 M^4 + 6 \big\{197 + L (L + 1)\big(11 L (L + 1)-98\big)\big\} M^2 \Theta 
\nonumber \\
& ~~~~ + (L^2+L-2)^2\big(12 + 5 L (L + 1) (L^2+L-5)\big) \Theta^2\big] M \Theta \\
\mathcal D_7 &= 16  \big[-54 \big(9 L (L + 1)-38 \big) M^4 + 9 (L -1) (L + 2) (L^2 + L -3)\big\{2 L (L + 1)
\nonumber \\
& ~~~~   -19\big\} M^2 \Theta + (L^2 + L -2)^3 (L^4 + 2L^3 - L + 1) \Theta^2\big] \Theta \\
\mathcal D_8 &= 864 M^5 - 432 (L^2 + L - 7) (3 L (L + 1) - 8) M^3 \Theta - 8 (L^2 + L - 2)^2 \times
\nonumber \\
& ~~~~ (32 L (L + 1) - 73) M \Theta^2 \\
\mathcal D_9 &= 432 (L^2 + L - 4) M^4 - 8 (L - 1) (L + 2) \big[538 + L (L + 1)\times   
\nonumber \\
& ~~~~ \big( 13 L (1 + L) - 187\big)\big] M^2 \Theta - 4 (L^2 + L - 5 ) (L^2 + L - 2 )^3 \Theta^2 \\
\mathcal D_{10}&= 72 \big[L (L + 1) (L^2 + L - 10) + 19\big] M^3 - 4 (L^2 + L - 2)^2 \big[L(L + 1) \times
\nonumber \\
& ~~~~ (L^2 + L - 21) + 92\big] M \Theta \\ 
\mathcal D_{11}&= 4 (L-1) (L + 2) \big[L (L + 1) (L^2 + L - 22) + 67  \big] M^2 
\nonumber \\
& ~~~~ + 2 (L^2 + L - 6) (L^2 + L - 2)^3 \Theta \\
\mathcal D_{12}&= -2 (L^2 + L - 2)^2 \big(2 L (L + 1) - 13 \big) M \\
\mathcal D_{13}&= (L^2 + L - 2)^3
\end{align}
and 
\begin{align}
\mathcal S(r)& = 192 \big[ 17280 M^6 + 576 \big(23 L (L + 1)-55 \big) M^5 r_0 + 144 (L -1) (L + 2) \nonumber \\ \nonumber
& ~~~~ \big(28 L (1 + L)-83\big) M^4 r_{0}^2 + 48 (L^2 + L-2)^2 \big(13 L (L + 1)-49\big) M^3 r_{0}^3 + \\ \nonumber
& ~~~~ 8 (L^2+L-2)^3 \big(7 L (L + 1)-31 \big) M^2 r_{0}^4 + 4 (L^2+L -3) (L^2+L-2)^4 M r_{0}^5 - \\
& ~~~~   L(L + 1) (L^2+L-2)^4 r_{0}^6\big] \Theta^4
\end{align}
In the next section, we will calculate the QNM frequency of the potential of eqn.~\eqref{even_th} and this frequency will be denote by $\omega^{\Theta}$.


\section{Numerical results for QNM}
\label{numer_qnm}

To estimate the QNMs for the potential of eqn.~\eqref{even_th} we will use the 6th order WKB formula (see eqn.~\eqref{6wkb} of Sec.~\ref{wkbkono}). We will also compare our results with that obtained from 3rd, 4th and 5th order formula in the WKB method. In order to calculate derivatives with respect to the coordinate $r_{\star}$ in the WKB formula (\ref{6wkb}) we proceed as follows: $\frac{dV}{dr_{\star}}=dV/dr \times dr/{dr_{\star}}$, and so on. Now plugging eqn.~\eqref{new_sph} into eqn.~\eqref{co_rst} and with the help of eqn.~\eqref{6wkb} the QNM frequencies can be determined. One can guess how complicated the form of QNM frequencies will be. That's why we will take care of numerical estimates of QNM frequencies.

In Table~\ref{tab_gwv}, we summarize the numerical values of the QNM frequencies (both Re${[\omega^{\Theta}}]$ and Im$[{\omega^{\Theta}}]$) of even parity gravitational perturbations for different orders of WKB formula with $L=2$ and $L=3$. This table shows how the $\Theta$ parameter affects the values of QNM frequencies. One can see that the 6th order WKB keeps significant effect to QNM frequencies compared to 3rd order WKB. There do not exist much differences in values of QNM frequencies between Schwarzschiled case and 3rd order WKB. As $\Theta$ increases, the 6th order WKB shows that the value of QNM decreases for $\Theta\lesssim 0.1$.   
\begin{table}[htb]
	\begin{center}
		\begin{tabular}{l|c|c|c|c|c}
			\hline
			\hline
			\bf L & \bf n & \bf Even & $\Theta$ & \bf Even & \bf Even  \\
			&  & \bf modes   & & \bf modes &  \bf modes \\
			\cline{5-6}
			&  & (Schwarzschild)  & & (Smeared matter) & (Smeared matter) \\
			&  &  & & 3rd order WKB & 6th order WKB \\
			\hline\hline
			2 & 0 &  0.373012  - i\,0.0891091  & 0.08 & 0.373014 - i\,0.0891205 & 0.37182 - i\,0.0899359 \\
			& &   & 0.1 & 0.37302 - i\,0.089306 & 0.357152  - i\,0.0907511   \\
			& &   & 0.12 & 0.372606 - i\,0.0890164 & 0.567636  - i\,0.0325815 \\
			\cline{2-6} 
			& 1 & 0.345164  - i\,0.274642 &  0.08  & 0.3452 - i\,0.274693 & 0.330056  - i\,0.293657 \\
			&  &   & 0.1 & 0.345623 - i\,0.275472 & 0.299666  - i\,0.272728 \\
			&  &   & 0.12 & 0.34263 - i\,0.273851 & 1.43322  + i\,0.0212183  \\
			\hline 
			3 & 0 &  0.599264  - i\,0.0927278 & 0.08 & 0.599017 - i\,0.0918392 & 0.599185  - i\,0.0918364 \\
			& &   & 0.1 & 0.599015 - i\,0.0918665 & 0.598298 - i\,0.0920111  \\
			& &   & 0.12 & 0.598948 - i\,0.091886 & 0.603859 - i\,0.089485 \\
			\cline{2-6} 
			& 1 &  0.58235 - i\,0.281404 & 0.08 & 0.580251 - i\,0.28018 & 0.580437  - i\,0.280231 \\
			&  &   & 0.1 & 0.580286 - i\,0.280322 & 0.574677 - i\,0.28266   \\
			&  &   & 0.12 & 0.57998 - i\,0.280373 & 0.626946 - i\,0.244375  \\
			\cline{2-6} 
			& 2 &  0.553187 - i\,0.476681 & 0.08 & 0.549744 - i\,0.477934  & 0.546845 - i\,0.481195 \\
			&  &   & 0.1 & 0.549963 - i\,0.478292 & 0.529764 - i\,0.49198 \\                             
			&  &   & 0.12 & 0.549149 - i\,0.478385 & 0.773037 - i\,0.28019 \\
			\hline
		\end{tabular} \vspace{0.3cm}
	\end{center}
	\caption{Comparison between the QNM frequencies for the gravitational perturbation of Schwarzschild spacetime 
		and spherically symmetric spacetime 
		with smeared matter source.
	}
	\label{tab_gwv}
\end{table}

\begin{figure}[t!]
	\centering
	\includegraphics[width=7.84cm]{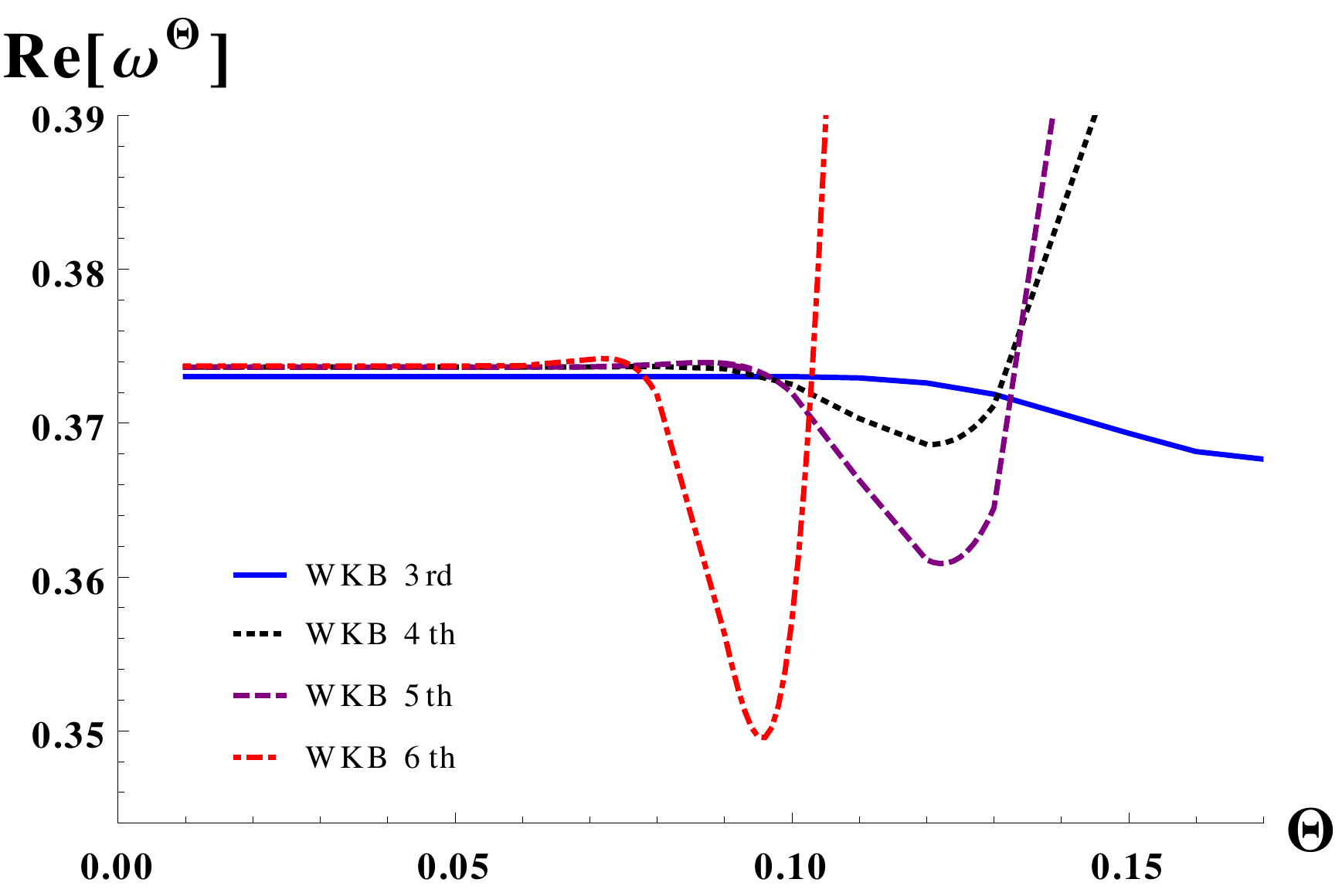}
	~
	\includegraphics[width=7.84cm]{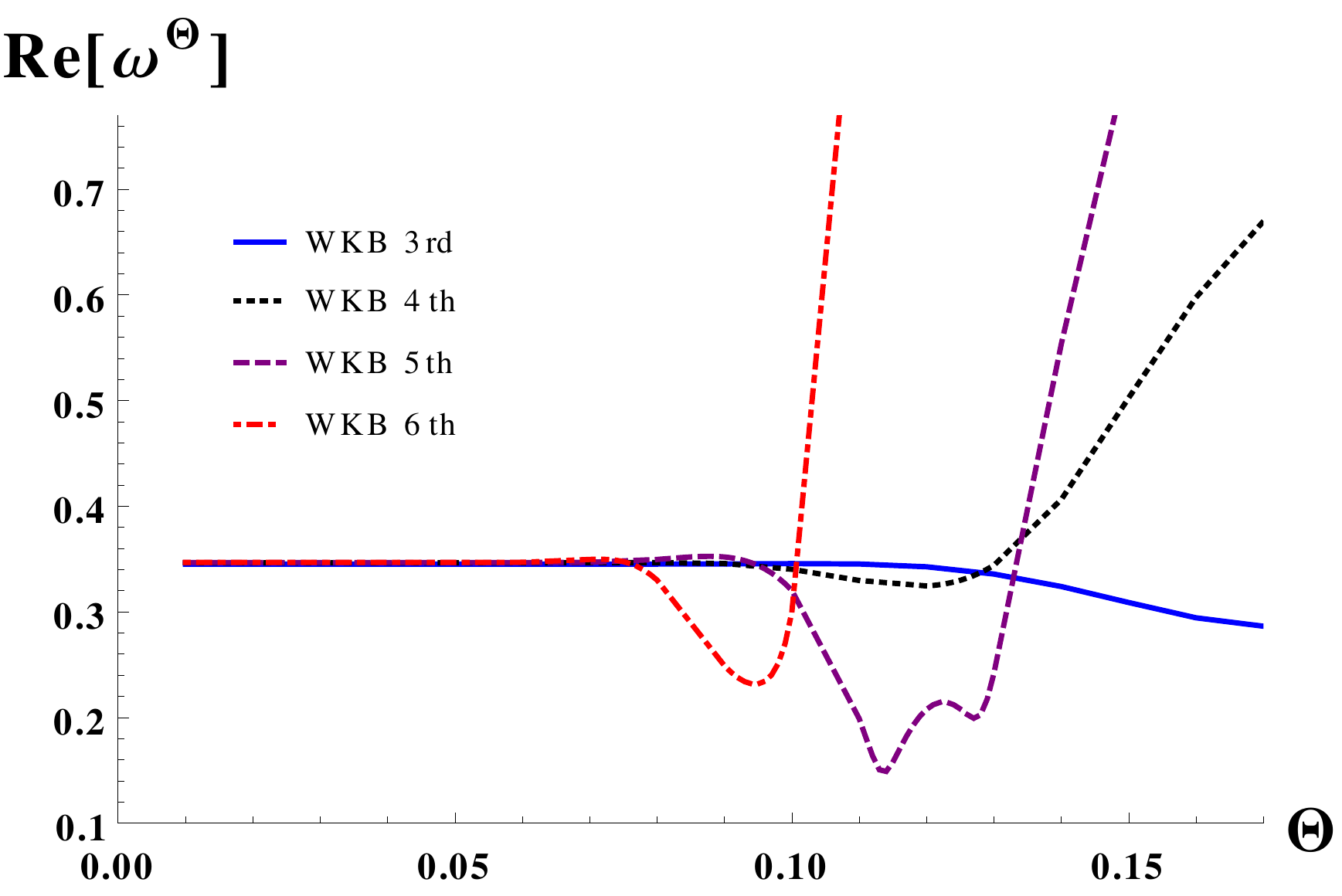}
	\caption{Plot showing the variation of the real part of QNM frequency $\omega^{\Theta}$, with increasing value 
		of the parameter $\Theta$. For the above plot we have considered $L=2,n=0$ (left) and $L=2,n=1$ (right) mode for different orders of the WKB approximation formula.  }
	\label{l2} 
\end{figure}

\begin{figure}[t!]
    \centering
    \includegraphics[width=7.84cm]{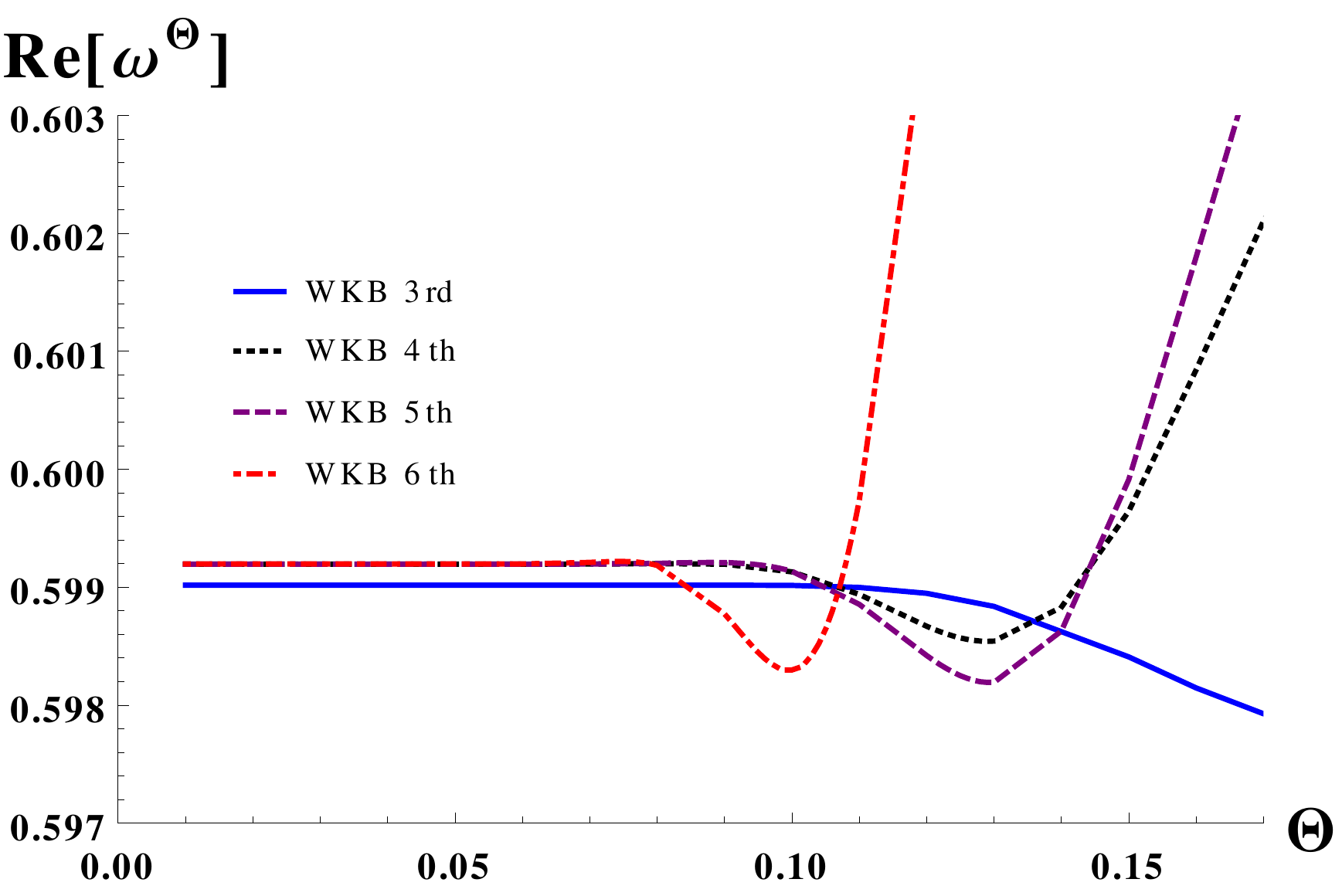}
~
\includegraphics[width=7.84cm]{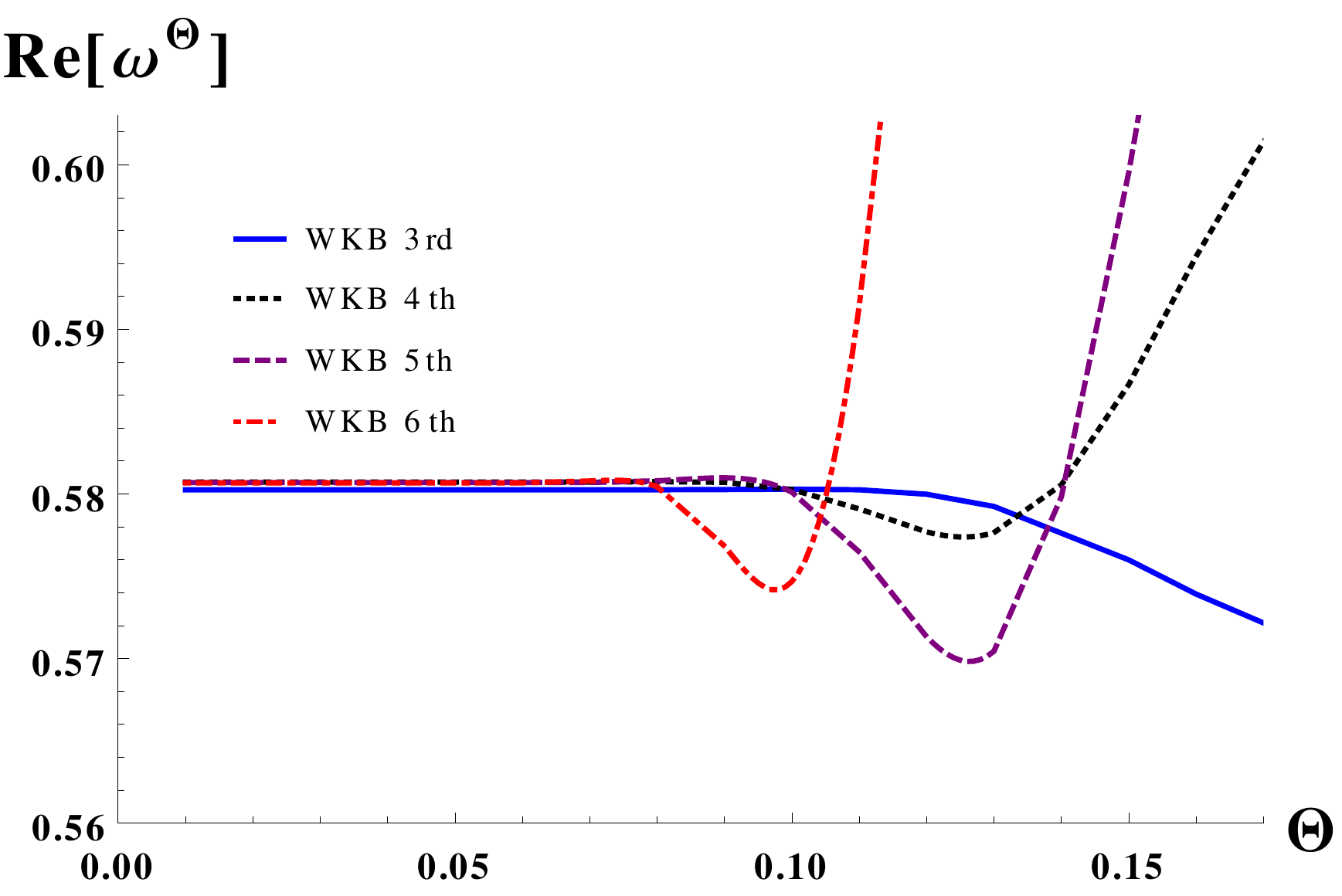}
~
\includegraphics[width=7.84cm]{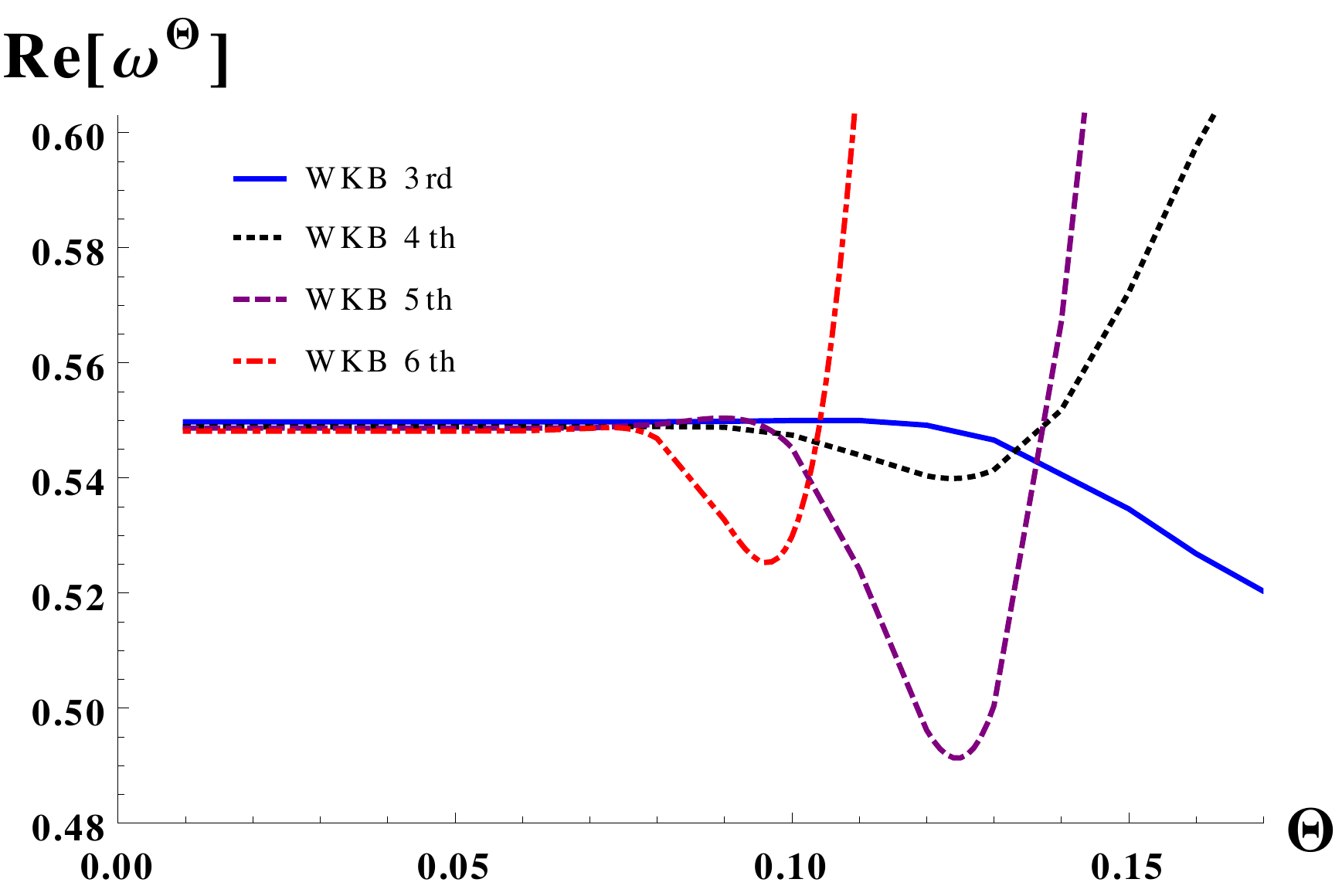}

             \caption{Plot showing the variation of the real part of QNM frequency $\omega^{\Theta}$, with increasing value 
of the parameter $\Theta$. For the above plot we have considered $L=3, n=0$ (topleft), $L=3, n=1$ (topright) and $L=3, n=2$ (bottom) mode for different orders of the WKB approximation formula.  } 
\label{l3}
\end{figure}

Fig.~\ref{l2} shows the variations in the real part of QNM frequencies for $ L=2$ $(n=0, 1) $ even parity gravitational perturbation as a function of the smearing length parameter $\Theta$. Here different colors correspond to different order approximations in the WKB formula, used to numerically compute the complex frequencies. We can clearly see from this plot that with the 3rd order WKB formula, Re${[\omega^{\Theta}}]$ remains almost constant for  $0\lesssim\Theta\lesssim 0.12$, and after that it decreases very slowly. But as we approach towards higher orders in the WKB series, constancy in Re${[\omega^{\Theta}}]$ is gradually shifted to lower values of $\Theta$ --- \emph{e.g.}  for the 4th and 5th order WKB formula, Re${[\omega^{\Theta}}]$ is roughly constant upto the regime $0\lesssim\Theta\lesssim 0.1$ and so is the case for the 6th order WKB formula upto $0\lesssim\Theta\lesssim 0.08$. 
For $L=3$ even mode, we found a similar behaviour in Re${[\omega^{\Theta}}]$ as can be seen from Fig.~\ref{l3} and we have checked that this result is true for other higher order even modes also. 
However, the spurious oscillations that show up in Re${[\omega^{\Theta}}]$ for $\Theta\gtrsim 0.12$ (in the 4th and 5th order WKB formula) and for $\Theta\gtrsim 0.1$ (in the 6th order WKB formula) implies that the corresponding order of the perturbation in WKB approximation scheme is valid upto that particular value of the model parameter. It seams that WKB approximation breaks down for higher values of the order parameter. This is because for higher values of $\Theta$ our assumption $r^2>> 4\Theta$ is no longer valid. Hence our model parameter $\Theta$ has an approximate  bound depending on the order of the WKB formula. Thus the accuracy of the bound gets stronger with increasing the orders of the WKB formula. This result is also consistent with our earlier work for odd parity perturbations \cite{Das:2018fzc}, where we showed the appreciable modifications of the 3rd order result by \cite{Liang:2018nmr,Liang:2018uyk} (see also \cite{Konoplya:2003ii} why orders higher than 6 are not feasible in this framework). 

\subsection{Isospectrality limit}
\label{limit_isospec}

\begin{figure}[t!]
    \centering
    \includegraphics[width=7.84cm]{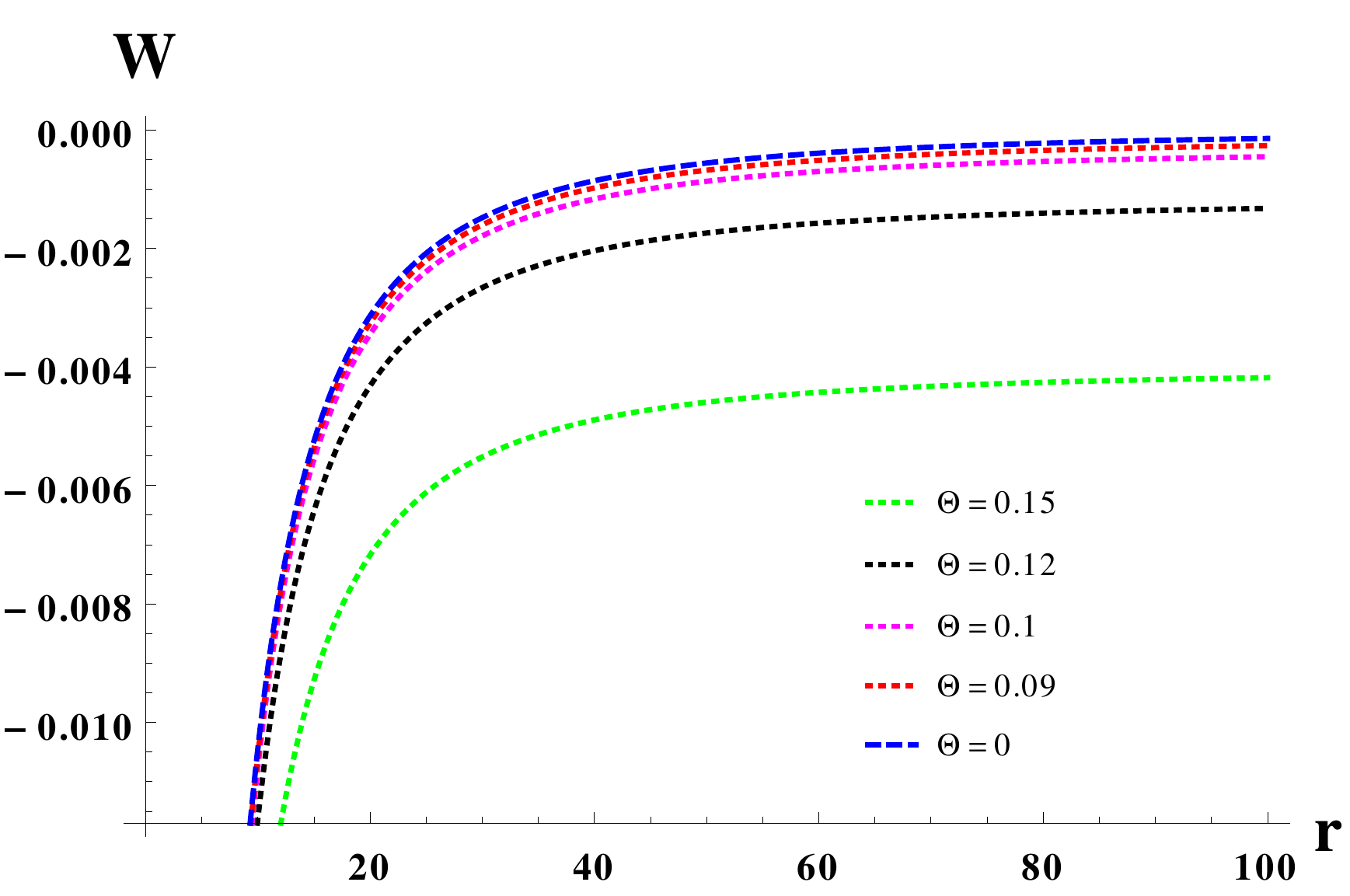}
\caption{Plot showing the variation of the function $W$ for different choices of the parameter $\Theta$ (taking $L=2$ and $M=1$)  }
\label{plot_iso} 
\end{figure}

It is a well established that, for spherically symmetric Schwarzschild like BH in an asymptotically flat spacetime, the odd and even parity gravitational perturbations yield the same spectra for the QNM frequencies \cite{Chandrasekhar:1985kt}. This is due to the fact that both the odd and even parity potentials can be expressed in terms of a common function $W$ as
\begin{align}
V_{odd} &= W^2 + \frac{dW}{dr_{\star}} + \beta, 
\label{wo} \\  
V_{even} &=  W^2 - \frac{dW}{dr_{\star}} + \beta, 
\label{we}
\end{align}
where $\beta$ is some function of the parameters $L$ and $M$. We would like to see whether a similar result holds for the spacetime of eqn.~\eqref{qg_sphe} with Gaussian source. In \cite{Das:2018fzc}, we obtained the gravitational odd parity potential of this spacetime which is of the form
\begin{align}
V_{odd}^{\Theta}(r) &= V_{odd} (r) + \frac{2M}{\sqrt{\pi\Theta}}e^{-r^2/{4\Theta}}\bigg[ \frac{l(l+1)}{r^2}-\frac{6M}{r^3}
+\bigg( \frac{1}{2\Theta} + \frac{2}{r^2} \bigg) \bigg(1-\frac{2M}{r}\bigg) \bigg] 
\label{odd_th} \\
\text{where,} \quad V_{odd}(r) & = \big(1-\frac{2M}{r}\big)\big(\frac{L(L+1)}{r^2} - \frac{6M}{r^3}\big)
\end{align}
The expression for the corresponding even mode potential is given in eqn.~\eqref{even_th}. Now subtracting eqn.~\eqref{we} from eqn.~\eqref{wo} and with the help of eqns.~\eqref{co_rst},\eqref{even_th} and \eqref{odd_th} we numerically compute the function $W$. The behaviour of $W$ has been shown in Fig.~\ref{plot_iso}. 
In this plot the dashed blue curve corresponds to the normal case (\emph{i.e.} the Schwarzschild BH or, $\Theta=0$) when the isospectrality exists between odd and even mode perturbations. 
 Other colored dotted lines are due to different choices of $\Theta$ parameter. We find that there is a similar limiting value, approximately $\Theta=\Theta_0\sim 0.1$, beyond which $W$ deviates significantly from its behaviour at $\Theta=0$.  
In fact, this limiting value of $\Theta$ was also obtained earlier  when the WKB approximation formula for the QNM frequencies induces oscillations in Re${[\omega^{\Theta}}]$ (see Figs.~\ref{l2} and \ref{l3}) and thereby making the results reliable upto $\Theta\sim\Theta_0$ . Therefore, for $\Theta\lesssim\Theta_0$, the 6th order WKB formula yields accurate predictions for the even mode perturbations and the potentials of opposite parities do share the same spectra in their associated QNM frequencies.

\section{Conclusion}
\label{conl}

In this work, we have studied the effects of minimal-spread matter distribution on the even mode gravitational perturbations of spherically symmetric spacetimes and computed the QNM frequencies. Previously, we have computed this study for odd parity perturbations of this same geometry. In general, for a spherically symmetric spacetime in four dimensions, the metric perturbations have both odd and even parity components. Here, we have derived the general form of the potential function governing the even modes of the metric perturbation. This result is new in the literature. In contrast to the usual Schwarschild BH, here the source term is replaced by a smeared matter distribution of the Gaussian type. As a result, the potential for even modes now involves a new length scale. The associated QNM frequencies of these modes are numerically estimated using the well known WKB formula. Further, we have made a detailed comparison amongst the various orders in the WKB approximation results for the frequency spectrum of even modes. We found that the WKB approximation is valid upto an approximate value of the smeared order parameter $\Theta$ within which the usual isospectrality between the perturbations of opposite parity, also holds good. However, beyond this limit they no longer share the same spectra. This work can be evaluated for other spherically symmetric geometries. It will be quite interesting to study the above theories for rotating objects.

{\bf{Acknowledgement:}} We thank specially to Professor Roman Konoplya for sending us the necessary code for computation. Kumar Das is supported by a Post-doctoral fellowship from S. N. Bose National Centre For Basic Sciences. Funds of Souvik Pramanik corresponds to the Science Education and Research Board, India.


\end{document}